\documentclass[aip,jap,reprint,superscriptaddress,amsmath,amssymb]{revtex4-1}
\usepackage{epsfig,psfrag}
\usepackage{dcolumn}
\usepackage{bm}
\usepackage{graphicx}
\usepackage{color}

\begin{document}
\title{Negative free carrier absorption in terahertz quantum cascade lasers}

\author{C. Ndebeka-Bandou}\email{ncamille@phys.ethz.ch}
\author{M. R\"osch}
\author{K. Ohtani}
\author{M. Beck}
\author{J. Faist}
\affiliation{Institute for Quantum Electronics, ETH Zurich, Auguste-Piccard-Hof 1, 8093 Zurich, Switzerland}

\begin{abstract}
We analyze the peculiar case where the free carrier absorption arising from LO phonon absorption-assisted transitions becomes negative and therefore turns into a gain source for quantum cascade lasers. Such an additional source of gain exists when the ratio between the electronic and the lattice temperatures is larger than one, a condition that is usually fulfilled in quantum cascade lasers. We find a gain of few cm$^{-1}$'s at 200~K. We report the development of a terahertz quantum cascade laser operating in the negative free carrier absorption regime. 
\end{abstract}


\maketitle

Quantum Cascade Lasers (QCLs) are unipolar devices that can achieve a stimulated  emission from the mid-infrared (IR) to the terahertz (THz) range \cite{faist1994,kohler2002,FaistBook2013}. Their high optical output power, compact size and operating frequency range make them very promising for fundamental research and industrial applications \cite{dean2014,yao2012}. Whereas mid-IR QCLs are already commercialized and operate at room temperature, no lasing THz QCLs has ever been reported above 200~K \cite{fathololoumi2012,sirtori2013} and their optimization is still an ongoing topic. There is room for improvement that includes the design\cite{kumar2014} and the material \cite{deutsch2010} optimization but also the development of advanced modellings to provide an accurate understanding of the transport and the loss/gain processes \cite{wacker2013,jirauschek2014}. Among the detrimental effects to the lasing action, the Free Carrier Absorption (FCA) is known to be an optical loss mechanism in QCLs \cite{vurgaftman1999,faist2007,wacker2011}. This process results in the reabsorption of the laser photons by the free carriers, in particular those that are located on the upper laser state. FCA arises from oblique  intra- and inter-subband transitions (in the $\vec{k}$ space) that are activated by the static scatterers and the phonons of the structure \cite{walukiewicz1979}. Recently, a quantum model of FCA in QC structures has been proposed \cite{carosella2012,iop2014}, thereby describing the mechanism as a three-body collision and focusing the attention on the loss magnitude and the respective contributions of the different scattering sources. It has been demonstrated that the oblique inter-subband transitions are the most efficient for photon reabsorption and that transitions activated by longitudinal optical (LO) phonon absorption/emission  dominate at high temperature. Moreover, it has been pointed out that, under specific conditions, FCA assisted by LO phonon absorption leads to a negative absorption coefficient and therefore provides gain to the device. 

In this paper, we report the design of a THz QCL active region based on this negative FCA principle.
Following the perturbative approach of Carosella \textit{et al.} \cite{carosella2012}, we first compute the FCA coefficient associated with inter-subband FCA assisted by LO phonon absorption and discuss the condition that the system has to fulfill to exhibit a negative FCA. Starting from this condition, we will discuss the different requirements for the design of a THz QCL active region operating in the negative FCA regime. Finally, we present our designed structure and its experimental characterization.

\begin{figure}
 \includegraphics{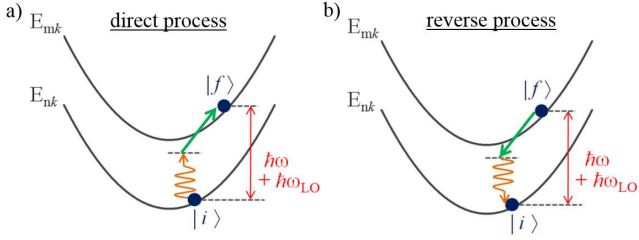}
 \caption{(Color online) Scheme of the FCA mechanism assisted by LO phonon absorption. The oblique inter-subband transition $|n\vec{k}\rangle\rightarrow|m\vec{k'}\rangle$ is considered where $E_{n,\vec{k}}$ is the dispersion of the upper lasing subband. a) Direct process corresponding to a photon absorption assisted by a LO phonon absorption. b) Reverse process corresponding to a photon emission assisted by a LO phonon emission. In both cases $E_{m\vec{k'}}-E_{n\vec{k}}=\hbar\omega+\hbar\omega_\mathrm{LO}$. }
\label{fig1}
\end{figure}

Among the different scattering sources in heterostructures, the LO phonon scattering is known to be of relevance and dominates at high temperature. To describe the inter-subband FCA assisted by LO phonon absorption, we consider an electron initially placed on the upper laser subband $n$ of the QCL active region. This electron is in the state $|i\rangle=|n,\vec{k}\rangle$ where $\langle \vec{\rho},z|n,\vec{k}\rangle=\frac{1}{\sqrt{S}}e^{i\vec{k}\cdot\vec{\rho}}\chi_n(z)$
with an energy $\varepsilon_{n\vec{k}}=E_n+\frac{\hbar^2k^2}{2m^*}$.  
$z$ is the growth direction, $\vec{\rho}$ refers to the in-plane position, $S$ is the sample area and $m^*$ is the electronic effective mass. $\vec{k}$ denotes the two-dimensional wavevector and $\chi_n(z)$ is the localized wavefunction for the bound motion in the $z$ direction.
In the presence of the laser mode, the FCA process can be schematically decomposed into two steps: i) the electron absorbs the laser photon of energy $\hbar\omega$ and  ii) absorbs a LO phonon of energy $\hbar\omega_\mathrm{LO}$ and is finally scattered to the state $|f\rangle=|m,\vec{k'}\rangle$ in the upper subband $m$ (see Fig.~\ref{fig1}a).  This process corresponds to the absorption of both the laser photon and the LO phonon and will be referred-to as the direct process. However, to compute the net FCA coefficient, we shall consider the reverse process as well, namely the inter-subband oblique transition that brings the electron from the state $|f\rangle$ to the upper laser state $|i\rangle$ by a photon emission and a LO phonon emission (see Fig.~\ref{fig1}b). Thus, considering the contribution of the absorption ($\alpha_\mathrm{LO\ abs}^\mathrm{direct}$) and its reversed  process (the stimulated emission, $\alpha_\mathrm{LO\ abs}^\mathrm{reverse}$), the net FCA coefficient is readily given by $\alpha^\mathrm{net}_\mathrm{LO\ abs}(\omega)=\alpha_\mathrm{LO\ abs}^\mathrm{direct}(\omega)-\alpha_\mathrm{LO\ abs}^\mathrm{reverse}(\omega)$.
Assuming an ideal overlap between the active region and the photon modes (necessarily $z$-polarized in QCLs) and following the perturbative approach of Ref.~\onlinecite{carosella2012}, one finds:
\begin{multline}
\alpha_\mathrm{LO\ abs}^\mathrm{net}(\omega)=\frac{2\pi(e/m^*)^2}{\omega\varepsilon_0cn_rLS}\sum_{\vec{k},\vec{k'},\vec{q}}\delta(E_{m\vec{k'}}-E_{n\vec{k}}-\hbar\omega-\hbar\omega_\mathrm{LO})\\
\times \left[  f_{n\vec{k}}(1-f_{m\vec{k'}})|\langle \Psi_{n\vec{k}},N_q |p_z|\Psi_{m\vec{k'}},N_q-1\rangle|^2 \right.\\\left.-f_{m\vec{k'}}(1-f_{n\vec{k}})  \right]
|\langle \Psi_{m\vec{k'}},N_q-1 |p_z|\Psi_{n\vec{k}},N_q\rangle|^2,
 \label{alpha_1}
\end{multline}
where $n_r$ the refractive index, $\varepsilon_0$ the vacuum dielectric constant and $c$ the speed of light. Note that the FCA coefficient is inversely proportional to $L$, the length of one QCL period.
The $\delta$-function in (\ref{alpha_1}) refers to the energy conservation $E_{m\vec{k'}}-E_{n\vec{k}}=\hbar\omega+\hbar\omega_\mathrm{LO}$ for the processes described in Fig.~\ref{fig1}.  The terms in the second and third line of (\ref{alpha_1}) contain the distribution functions $f_{n,\vec{k}}$ and $f_{m,\vec{k'}}$ of the electronic states $|n\vec{k}\rangle$ and $|m,\vec{k'}\rangle$ respectively. Due to the low carrier densities (typically few $10^{10}$~cm$^{-2}$ in THz QCLs \cite{kohler2002,FaistBook2013}), we assume Boltzmann distributions for thermalized subbands at a temperature $T_e$.  $N_q$ is the number of LO phonons of energy $\hbar\omega_q\approx\hbar\omega_\mathrm{LO}$ ($\hbar\omega_\mathrm{LO}=36$~meV in GaAs) and is given by the Bose-Einstein factor at the lattice temperature $T_L$. $\vec{q}$ denotes the three-dimensional phonon wavevector. The dipole matrix elements in (\ref{alpha_1}) account for the coupling between the laser mode and the perturbed mixed electron-LO phonon states $|\Psi_{n\vec{k}},N_q\rangle$ and $|\Psi_{m\vec{k'}},N_q-1\rangle$ that contain either $N_q$ or $N_q-1$ phonons  (see Ref.~\onlinecite{carosella2012} for the full derivation). 

It is important to recall that the absorption contribution is proportional to the number of phonons $N_q$ while the contribution of the stimulated emission is proportional to $N_q+1$. This difference of prefactors is in fact the origin of the negative FCA. In other words, this effect stems from the fact that the scatterers are bosons with a temperature-dependent distribution. This feature cannot be found while considering FCA assisted by elastic scatterers. 

\begin{figure}
 \includegraphics{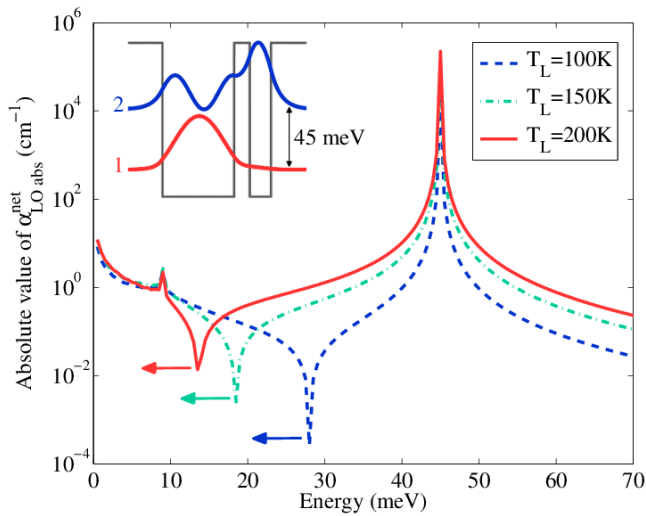}
 \caption{(Color online) Absolute value of the FCA coefficient $\alpha^\mathrm{net}_\mathrm{LO\ abs}$ versus the photon energy $\hbar\omega$ for a lattice temperature $T_L=100$~K (blue dashed line), $T_L=150$~K (green dashed-dotted line) and $T_L=200$~K (red solid line). The electronic temperature is set to $T_e=T_L+80$~K. The arrows correspond to the energy range where $\alpha_\mathrm{LO\ abs}^\mathrm{net}<0$.  The calculation has been done for a 11.5/2.5/3.4~nm GaAs/Al$_{0.15}$Ga$_{0.85}$As double quantum well. The oblique $|1\vec{k}\rangle\rightarrow|1\vec{k'}\rangle$ and $|1\vec{k}\rangle\rightarrow|2\vec{k'}\rangle$ transitions contribute to the FCA coefficient. The inset shows the band diagram of the structure and the squared moduli of the states involved in the FCA process.  $n_e=3\times10^{10}$~cm$^{-2}$. $E_{mn}=45$~meV.
}
\label{fig2}
\end{figure}

From (\ref{alpha_1}) follows that if the reverse process is dominant, the FCA coefficient reaches a negative value and turns into gain for the device. This gain differs from the scattering-assisted Bloch gain \cite{willenberg2003} since it corresponds to a photon emission at exactly the lasing frequency. Moreover, while the scattering-assisted Bloch gain dominates when the population of the upper and lower lasing states are equal (weak population inversion), the negative FCA requires to maintain an efficient population inversion to produce the laser photon which activates the FCA transitions.  By calculating explicitly (\ref{alpha_1}), one finally obtains:
%
\begin{multline}
 \alpha^\mathrm{net}_\mathrm{LO\ abs}(\omega)=\frac{e^4n_e\omega_\mathrm{LO}N_q}{16\pi \varepsilon_0^2\varepsilon_pcn_rm^*L\hbar\omega}\frac{|\langle n |p_z|m\rangle|^2}{(\hbar\omega-E_m+E_n)^2}
 \\\times \left[ (1-e^{-\beta_e\hbar\omega}e^{(\beta_L-\beta_e)\hbar\omega_\mathrm{LO}})I_\mathrm{abs}^{(nm)}
+
\frac{n_e\hbar^2\pi}{2m^*k_BT_e}\right.\\\left.\times(e^{-\beta_e\hbar\omega}e^{(\beta_L-\beta_e)\hbar\omega_\mathrm{LO}}-e^{-\beta_e(\hbar\omega+\hbar\omega_\mathrm{LO})})I'^{(nm)}_\mathrm{abs}
\right],
 \label{alpha_2}
\end{multline}
%
where $n_e$ denotes the carrier sheet density, $I^{(nm)}_\mathrm{abs}$ and $I'^{(nm)}_\mathrm{abs}$ are integrals explicitly given in Eq.~15 of Ref.~\onlinecite{carosella2012}. $\beta_{e,L}=(k_BT_{e,L})^{-1}$. $\varepsilon_p^{-1}=(\varepsilon_\infty^{-1}-\varepsilon_s^{-1})$ where $\varepsilon_\infty$ and $\varepsilon_s$ are the high-frequency and the static relative permittivities of the heterostructure respectively. From (\ref{alpha_2}), the condition for a negative FCA is derived:
\begin{equation}
\hbar\omega< \hbar\omega_\mathrm{LO}\left( \frac{T_e}{T_\mathrm{L}}-1 \right).
\label{negative_fca}
\end{equation}
A similar expression is obtained for intra-subband transitions \cite{carosella2012,iop2014}.  Eq.~(\ref{negative_fca}) shows that as soon as $T_e>T_L$ - a condition usually assumed in QCLs \cite{vitiello2006,patimisco2013} - the FCA coefficient becomes negative over a frequency range which is strongly controlled by the ratio of temperatures $T_e/T_L$.
As a representative example, 
Fig.~\ref{fig2} shows the absolute value of the FCA coefficient versus the photon energy at various temperatures. For this calculation, we have considered a double quantum well structure and we have assumed an electronic temperature $T_e=T_L+80$~K \cite{vitiello2006,patimisco2013}. The plots display several interesting features. We first identify the resonance at $\hbar\omega\approx45$~meV arising when the photon energy strictly equals the inter-subband spacing $E_{mn}$, in agreement with Eq.~\ref{alpha_2}. The small peak located at $\hbar\omega=E_{mn}-\hbar\omega_{LO}\approx9$~meV corresponds to the phonon replica for LO phonon absorption, i.e., where the electron-phonon interaction is enhanced due to the $1/q$-dependence of the Fr\"ohlich potential. Since the absolute value of $\alpha_\mathrm{LO\ abs}^\mathrm{net}$ is plotted in log scale, the drastic change of the curve variation corresponds to a  change of the sign of the absorption. This occurs at a photon energy of $\hbar\omega<28.8$~meV, $19.2$~meV and $14.4$~meV for $T_L=100$~K, $150$~K and $200$~K respectively (in Fig.~\ref{fig2}, the energy range of the negative FCA is marked by the arrows).
The energy range where the negative FCA manifests itself gets narrower when $T_L$ increases due to a decreasing ratio $T_e/T_L$ but remains suitable for THz operations at 200~K. We also note that the magnitude of the FCA coefficient increases with the temperature. This effect is mainly due to the increasing number of phonons in the structure. 

Furthermore, it is important to specify that while transitions assisted by LO phonon absorption, i.e., where $E_{m\vec{k'}}-E_{n\vec{k}}=\hbar\omega+\hbar\omega_\mathrm{LO}$, can lead to gain, such is not the case of transitions assisted by LO phonon emission for which $E_{m\vec{k'}}-E_{n\vec{k}}=\hbar\omega-\hbar\omega_\mathrm{LO}$ \cite{supplemental}.

In the following, we concentrate on the design of a THz QCL operating in the negative FCA regime. To engineer such a structure, several requirements are needed: i) the designed lasing energy must fulfill the condition (\ref{negative_fca}), ii) the negative FCA must overcome the other loss contributions, i.e., the positive FCA arising from the elastic scatterers and the other phonon-assisted processses (LO phonon emission and LA phonon absorption/emission). Moreover, it is possible to increase the magnitude of the negative FCA by taking benefit from the phonon replica where the electron-LO phonon wavevector exchange is minimum. The oscillator strength between the states $|n\rangle$ and $|m\rangle$ involved in the oblique transition can also be adjusted by a suitable design.

Using a one-dimensional Schr\"odinger-Poisson solver and the density matrix formalism \cite{FaistBook2013}, we have designed a four-well Al$_{0.15}$Ga$_{0.85}$As/GaAs active region for THz emission in the negative FCA regime.
\begin{figure}
 \includegraphics{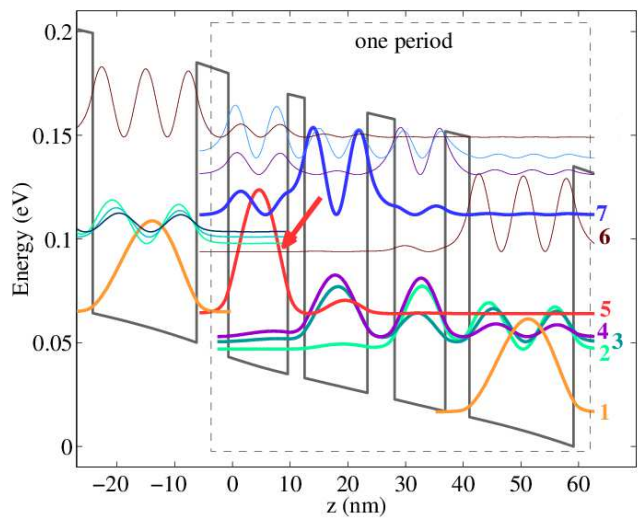}
 \caption{(Color online) Conduction band diagram of one period of the structure operating in the negative FCA regime at the design bias $F=7.2$~kV/cm. Starting from the injection barrier, the layer sequence is \textbf{5.5}/10.3/\textbf{2.9}/10.9/\textbf{4.7}/8.8/\textbf{4.2}/\underline{18.0}~nm where the thicknesses of the Al$_{0.15}$Ga$_{0.85}$As barriers are written in bold and the underlined one corresponds to the doped layer. The lasing action takes place between states 5 and state 4 while the negative FCA transition, indicated by the red arrow, occurs between states 5 and state 7.
}
\label{fig3}
\end{figure}
Figure~\ref{fig3} displays the conduction band diagram and the squared moduli of the relevant wavefunctions of the structure at the design bias $F=7.2$~kV/cm. The lasing action takes place between states $5$ and $4$ at an energy of 11.9~meV. The miniband formed of states $3$ and $2$ ensures the extraction of the carriers to the widest well for a fast depopulation by phonon emission. The negative FCA transition takes place between the upper laser state $5$ and state $7$ and the spacing $E_{75}$ is resonant with the phonon replica ($E_{75}\approx48$~meV). The widest well is uniformely doped to $2\times10^{16}$~cm$^{-3}$ per period corresponding to a carrier sheet density of $n_e=3.6\times10^{10}$~cm$^{-2}$. 

\begin{figure}
 \includegraphics{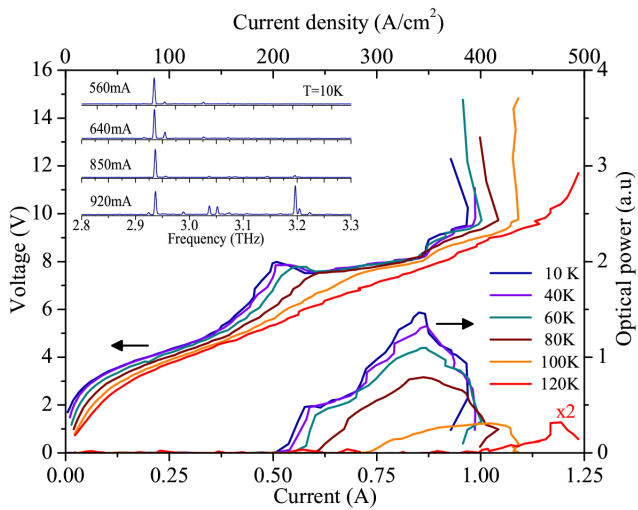}
 \caption{(Color online) Emitted light and bias-voltage versus current at various temperatures. The inset shows the emission spectrum of the device measured with a commercial FTIR (Bruker Vertex 80v) for different currents at 10~K. 
}
\label{fig5}
\end{figure}

The designed structure has been grown by molecular beam epitaxy to yield a 12.4~$\mu$m thick active region containing 180 periods. The active region was processed into a metal-metal waveguide and experimentally characterized. Figure~\ref{fig5} shows the light and bias-voltage versus current characteristics in pulsed mode at various temperatures. A lasing emission at the designed frequency was obtained at a threshold current density of $200$~A/cm$^2$ at $10$~K. The device reached a maximum operating temperature of $120$~K. 

\begin{figure}
 \includegraphics{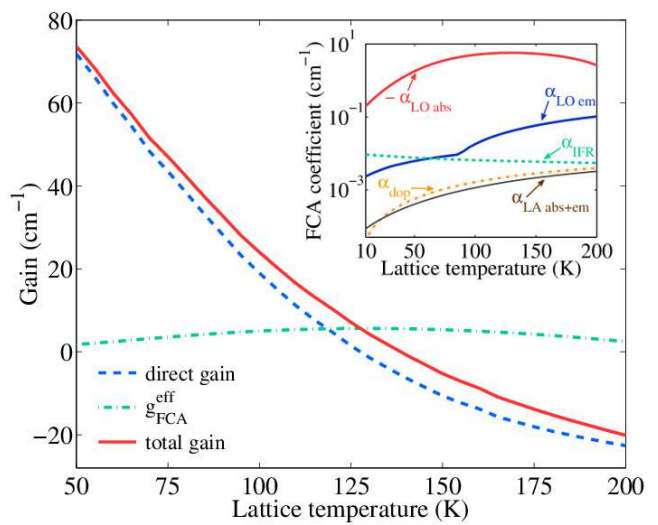}
 \caption{(Color online) Simulated gain  versus temperature in the designed 4-well QCL. The direct gain (blue dashed line) corresponds to the QCL gain proportional to the population inversion.  The effective negative FCA $g_\mathrm{FCA}^\mathrm{eff}$ (green dashed-dotted line) is equal to the sum of the different scattering contributions displayed in the inset as a function of the lattice temperature.  The total gain (red solid line) is the sum of the effective negative FCA and the direct gain. The inter-subband transition  $|5\vec{k}\rangle\rightarrow|7\vec{k'}\rangle$ as well as the intra-subband one ($|5\vec{k}\rangle\rightarrow|5\vec{k'}\rangle$) have been included in the calculation of the FCA coefficients showed in the inset. The electron temperature is set to $T_e=T_L+80$K. 
}
\label{fig6}
\end{figure}

Using Eq.~\ref{alpha_2} and following Refs.~\onlinecite{carosella2012,iop2014}, we have calculated the FCA coefficients associated with the scattering by interface roughness ($\alpha_\mathrm{IFR}$), dopants ($\alpha_\mathrm{dop}$), LO phonon ($\alpha_\mathrm{LO\ abs/em}$) and LA phonon ($\alpha_\mathrm{LA\ abs+em}$) emission/absorption. The effective FCA gain ($g_\mathrm{FCA}^\mathrm{eff}$) is given by the sum of all these contributions. Both the inter-subband $|5\vec{k}\rangle\rightarrow|7\vec{k'}\rangle$ and the intra-subband $|5\vec{k}\rangle\rightarrow|5\vec{k'}\rangle$ transition are included in the calculations for a full quantitative estimate. The results are shown in Fig.~\ref{fig6} versus the lattice temperature and at the lasing energy. In addition, we show the simulated direct gain - linearly dependent on the population inversion - as well as the total gain of the device. The negative contribution of the LO phonon absorption prevails over the other losses (positive FCA arising from the other scattering mechanisms, see inset of Fig.~\ref{fig6}). An additional gain of about 2-3~cm$^{-1}$ is found at high temperature. This value is not negligible compared to the actual optical gain in THz QCLs and could encourage high temperature operations but remains too small to be clearly distinguishable from the direct gain in the present device.

In conclusion, we have theoretically analyzed the negative FCA in THz QCLs. Such an additional source of gain arises from oblique intra- and inter-subband transitions assisted by LO phonon absorption and manifests itself over a specific energy range that depends on the ratio between the electronic and the lattice temperatures. We have designed, grown and characterized a four-well THz QCLs operating in the negative FCA regime. We have obtained a lasing emission up to 120~K and similar transport and optical characteristics as standard THz QCLs, indicating that the insertion of the FCA transition in the design does not jeopardize the lasing action. Despite the difficulties to benchmark the negative FCA contribution experimentally, we have theroretically shown that 2-3 cm$^{-1}$'s can be reached at high temperature. Thus, combining the negative FCA process and an optimization of the injection/extraction barrier thicknesses \cite{dupont2010} could enable to achieve higher temperature operations, espacially in shorter active regions such as three- or two-well structures where the negative FCA is expected to be larger. Moreover, further additional experiments, such as temperature-dependent gain  measurements, may reveal the existence of FCA-assisted gain processes.

\begin{acknowledgements}
C-NB wants to thank the support from the ERC under the project MUSiC and Profs G. Bastard and A. Wacker for valuable discussions. MR acknowledges financial support by the EU research project TERACOMB (call identifier FP7-ICT-2011-C, project no. 296500). The experimental work has been supported by ETH Zurich with the FIRST cleanroom facility.
\end{acknowledgements}


%

\end{document}